\documentclass[a4paper, 10pt, conference]{ieeeconf}      % Use this line for a4 paper
\usepackage{amsmath}
\usepackage{lmodern}
\usepackage{graphicx}
\usepackage{algorithm}
\usepackage{algorithmic}
\usepackage{times}
\IEEEoverridecommandlockouts                              % This command is only needed if 
\title{\LARGE \bf
A Collision-Free Path Planning Algorithm for Unmanned Aerial Vehicle Delivery
}

\author{Ziji Shi$^{1}$, Wee Keong Ng$^{2}$% <-this % stops a space
\thanks{$^{1}$Ziji Shi is a student of School of Computer Science and Engineering, 
        Nanyang Technological University, Singapore
        {\tt\small ZSHI005@e.ntu.edu.sg}}%
\thanks{$^{2}$Wee Keong Ng is a faculty with School of Computer Science and Engineering, 
        Nanyang Technological University, Singapore
        {\tt\small awkng@ntu.edu.sg}}%
}

\begin{document}

\maketitle
\thispagestyle{empty}
\pagestyle{empty}

%%%%%%%%%%%%%%%%%%%%%%%%%%%%%%%%%%%%%%%%%%%%%%%%%%%%%%%%%%%%%%%%%%%%%%%%%%%%%%%%
\begin{abstract}

Path planning is important for the autonomy of Unmanned Aerial Vehicle (UAV), especially for scheduling UAV delivery. However, the operating environment of UAVs is usually dynamic. Without proper planning, collisions may happen when multiple UAVs are congested. There may also be temporary no-fly zone setup by authorities that makes some airspace unusable. Thus, proper pre-departure planning that avoid such conditions is needed. 
 
In this paper, we formulate this problem into a Constraint Satisfaction Problem to find a collision-free shortest path on a dynamic graph. We proposed a collision-free path planning algorithm that is based on A* algorithm. The main novelty is that we discovered a heuristic function that also considers waiting time. We show that the proposed algorithm is optimal since the heuristic is admissible. We then implemented this algorithm in simulation using Singapore's airspace structure. Our simulation exhibited desirable runtime performance. Using the proposed algorithm, we found that the percentage of collision-free routes decreases as number of requests per unit area increases, and this percentage drops significantly at boundary value. Our empirical analysis could aid the decision-making of no-fly zone policy and infrastructure of UAV delivery. 
\end{abstract}

%%%%%%%%%%%%%%%%%%%%%%%%%%%%%%%%%%%%%%%%%%%%%%%%%%%%%%%%%%%%%%%%%%%%%%%%%%%%%%%%
\section{INTRODUCTION}

UAV delivery is growing from a concept to a reality. In places where ground transportation is unavailable or congested, UAV delivery could be a fast and low-cost alternative. It could serve in infrequent use case like delivery of emergency relief supplies, or commercial delivery of packages for online shopping. In both use cases, path planning is important. Rescue teams want to get the needed aid delivered as soon as possible, while owners of commercial UAVs may wish to maximize the monetary benefits from their UAVs by taking shorter routes.

However, there are many realistic constraints in the path planning. One type of constraints is threats in the air, for instance, other aerial vehicles. Just like vehicles running on the road, UAVs may collide with each other in the air as well, and the chance of collision grows as the geographical density of UAVs increases. In densely populated area, this collision could be disastrous. Furthermore, in modern city with many skyscrapers, there is limited space for maneuvers, thus we must avoid dispatching too many UAVs in a small space in a short period of time. In the setting of commercial delivery, each UAV owner may possess hundreds or even thousands of UAVs, which increases the chances of collision. Therefore the owners should strive to avoid collision during path planning stage. 

Another type of constraints is the unavailability of air space due to no-fly zone or extreme weather in certain area. It may be worth noting that this no-fly zone should be temporary, otherwise it should have been removed from the usable airspace structure. Examples could be sports events or carnivals, where no-fly zone is setup temporarily for the safety of participants. It could also be periodical no-fly zones, like downtown or Central Business District where it is densely populated during daytime and emptier during the night.

In this paper, we propose an algorithm to output shortest, collision-free, flyable path based on A* algorithm. The main contribution is a heuristic function to calculate the penalty while waiting due to traffic control. We use a data structure called \textit{CurrentSchedule}for checking and updating the availability of airspace. We generated and simulated a large amount of delivery request with airspace structure in Singapore. Simulation shows that the proposed algorithm scales almost at linear with regard to the number of requests until some equilibrium points.

The rest of paper is organized as follows. Related works are presented in Section 2. Section 3 gives a formal formulation of problem, and Section 4 describes our algorithm followed by a discussion on its optimality. Section 5 details necessary information about implementation of the proposed method. Analysis of experimental results are conducted in Section 6, and finally, conclusion and future works of the study is depicted in Section 7.

%%%%%%%%%%%%%%%%%%%%%%%%%%%%%%%%%%%%%%%%%%%%%%%%%%%%%%%%%%%%%%%%%%%%%%%%%%%%%%%%
\section{RELATED WORK}

The UAV path planning algorithms generally fall under two broad categories: one is the offline path planning algorithm, which uses global information about the environment to generate optimal path; the other is the online path planning algorithm, which employs information perceived through sensors and reroute on-the-fly.

Online path finding algorithms is designed to deal with uncertainties and emergencies during the flight. \cite{c1} presents some common online algorithms, including potential field approach and particle swarm optimization (PSO). They have advantages and disadvantages: potential field approach are simple and straightforward to understand, but may easily fall into local optimum should no adjustment to algorithm is done. PSO is also intuitive, but it can also be too computationally expensive to reroute in real-time. Boivin et al. \cite{c8} proposed a predictive control scheme that use shared knowledge between nearby UAVs to find collision-free route, meanwhile considering the dynamics of UAV itself. 

A fundamental challenge to online algorithms is the modelling of uncertainties during the flight. To model the impact of derivation of trajectory to routing, Kim \cite{c9} proposed a probabilistic trajectory model in 3D space. In his model, multiple trajectories are proposed with different degree of derivations to the baseline trajectory. The probability of conflicts is calculated for each potential trajectory, and the one with lowest 

Off-line algorithms plan the path prior to departure. The aircraft are usually modelled as point mass moving in two dimensions with constant speed and upper-bounded turning rate. \cite{c2} uses Genetic Algorithm (GA) to find optimal flyable path through numerous iterations, with each iteration improving previous path by rerouting at waypoint. The problem with Genetic Algorithm is that runtime grows exponentially as number of iterations and flights grows. \cite{c3,c4} also uses GA, but with a parallel approach to speedup the calculation. \cite{c10} formulates vehicle dynamic model, collision avoidance constraint, and multiple waypoint constraint as mixed-integer linear program (MILP). One of its novelty is that it uses trigonometric functions to better approximate the radius constraint. However, it does not consider battery constraint, which is a limiting factor to UAV's delivery capability. In this paper, we will address both collision avoidance constraint and battery constraint. 

%%%%%%%%%%%%%%%%%%%%%%%%%%%%%%%%%%%%%%%%%%%%%%%%%%%%%%%%%%%%%%%%%%%%%%%%%%%%%%%%
%%%%%%%%%%%%%%%%                    TODO                             %%%%%%%%%%%
%%%%%%%%%%%%%%%%%%%%%%%%%%%%%%%%%%%%%%%%%%%%%%%%%%%%%%%%%%%%%%%%%%%%%%%%%%%%%%%%
Besides, there are other algorithms with emphasis on different aspects. 

There are also hybrid solutions where both offline and online combination are used. This is also often the common practice for commercial companies. 

In the use case of delivery, its is often that routing is on-demand, which means we are required to route a new request frequently, although it may be sub-optimal. Also, new path should fit into our existing schedules, which satisfies the collision-free and no-fly zone constraint.

Given we have global information regarding the the existing schedules of UAVs and their current locations, we can optimize the route through off-line planning. One natural way is using A* algorithm. A* algorithm \cite{c5} is a best-first, efficient and optimal search algorithm. It uses heuristics to guide the search, and therefore, reduce the number of nodes needed to be explored. Another advantage for using A* is that it can apply on both 2D and 3D space as long as the graph is connected. Zhang et al. \cite{c6} proposed an offline improved A* algorithm to deal with realistic constraints in UAV movement, including maximum moving angle, minimum route leg length, minimum flight height and maximum route length. Their algorithm can avoid collision with terrain while minimize the flight height. However, it only handles static graph, and can be computationally expensive. Zhang et al. proposed fixes by trimming some less-permissble nodes, or imposing some hard nodes that the route must follows, but doing so will lose some potentially optimal solutions, and thus lose its completeness and optimality.

%%%%%%%%%%%%%%%%%%%%%%%%%%%%%%%%%%%%%%%%%%%%%%%%%%%%%%%%%%%%%%%%%%%%%%%%%%%%%%%%
\section{PROBLEM FORMULATION}

We model UAV as point mass moving in two dimensions with a constant cruising speed and a maximum flying time. The fixed altitude is because air space is usually structured in layers \cite{c10, c11}. Therefore, our proposed collision-free dynamic routing problem is formulated as: 
\begin{quote}
    Given a set of existing flight schedules $M$, no-fly zone schedule $S$, a graph $G(V,E)$ consists of edges $E$ and vertices $V$, and a new request $R_i(o,d,t)$ to route from $o \in V$ to $d \in V$ at time $t$, we would like to find out if there is a viable new route from $o$ to $d$, meanwhile, it should not pass through any no-fly zone and existing flight. If such a solution exists, output the solution paths and time needed.
\end{quote}

%%%%%%%%%--------------------------------------------------------%%%%%%%%%
\subsection{Notations} 
We use two binary variables $U_{e}(u,e_i,t)$ and $U_{v}(u,v_j,t)$ to represent if aircraft is occupying an edge $e_i$ or vertex $v_j$ respectively. When $U_{e}(u,e_i,t) = 1$, it means "at time $t$, UAV $u$ is flying on edge $e_i$". Similarly, $U_{v}(u,v_j,t) = 1$ means "at time $t$, UAV $u$ is flying on node $v_j$".

Waiting penalty refers to the sum of expected waiting time for taking an edge or node ($edgePen(e,t)$ and $nodePen(v,t)$). It can be set as a constant number, or a function whose value depends on traffic density, time, and other factors like air space structure.

Thus, we use $P(u,t)$ to represent the waiting penalty at time $t$ for UAV $u$.

\begin{equation} \label{eq:1}
\begin{split}
P(u,t) & =  U_{e}(u,e_i,t)*edgePen(e_i,t)\\
        & + U_{v}(u,v_j,t)*nodePen(v_j,t)
\end{split}
\end{equation}

% \[ \begin{cases} 
%       0 & x\leq 0 \\
%       \frac{100-x}{100} & 0\leq x\leq 100 \\
%       0 & 100\leq x 
%   \end{cases}
% \]

%\sum_{i=1}^{E}\sum_{j=1}^{V}\sum_{t=1}^{T} U_{e}

%%%%%%%%%--------------------------------------------------------%%%%%%%%%
\subsection{Constraints}

There are three realistic constraints in this problem, namely, collision-free constraint, no-fly zone constraint, and maximum flyable time constraint (battery constraint): 

\subsubsection{Collision-free Constraint} 
Avoidance of collision is critical to proper routing for aircraft. Collision-free has two aspects of meaning:

\begin{equation}
    \begin{split}
    \forall v \in V, u_i,u_j & \in U \land u_i\ne u_j U(u_i,v,t_i) \land U(u_j,v,t_j)\\
                            &\rightarrow t_i \ne t_j 
    \end{split}
\end{equation}

    There should be no other UAVs occupying the same node at the same time. There are two types of nodes, connecting nodes and landable nodes. Connecting nodes are geometrical nodes that UAVs can make turns, thus, all nodes in the airspace graph are connecting nodes. Landable nodes are nodes where we have UAV dispatching stations to recharge battery, load and unload packages, and conduct maintenance. Each trip must start and end from landable nodes. 
    
\begin{equation}
    \begin{split}
    \forall e \in E, u_i,u_j & \in U \land u_i\ne u_j U(u_i,e,t_i) \land U(u_j,e,t_j)\\
                            &\rightarrow t_i \ne t_j 
    \end{split}
\end{equation}

    There should be no more than one UAVs occupying the same edge at the same time. This addresses the concern where two UAVs may collide with each other. \\
    
\subsubsection{No-fly Zone Constraint} 
No-fly zone (NFZ) is a designated area where flight in that area is prohibited for a period of time, but UAV may fly across it before and after the designated time. This information can be obtained prior to path planning. 
To avoid flying across no-fly zone, we pre-process the no-fly zone schedule and place some virtual UAVs on the "no-fly" edges and nodes. From equation \ref{eq:1}, the waiting penalty for using such edges will be prohibitively high. Thus, we can filter out those routes who will cross no-fly zone using A* algorithm.

Let $E_{nfz}$ be a set of edges in no-fly zone, $V_{nfz}$ be the set of nodes in no-fly zone, $T_{start}$ be the start time and $T_{end}$ be the end time, then we have:

\begin{equation} \label{nfz1}
    \begin{split}
    \forall u_i \in U,  &U(u_i,e,t_i)\\
        & \rightarrow e\not\in E_{nfz} \lor t_i < T_{start} \lor t_i > T_{end}
    \end{split}
\end{equation}
    
\begin{equation} \label{nfz2}
    \begin{split}
    \forall u_i \in U, &U(u_i,v,t_i)\\
        & \rightarrow v\not\in V_{nfz} \lor t_i < T_{start} \lor t_i > T_{end}
    \end{split}
\end{equation}

Equation \ref{nfz1} means UAV cannot cross any edge of no-fly zone from the start to the end of no-fly period. Similarly, equation \ref{nfz2} means UAV cannot hover on nodes that belong to no-fly zone.

\subsubsection{Maximum Flight Time Constraint}
Battery is a realistic constraint in UAV delivery. Because of it, we use time instead of distance as weight in our A* algorithm because battery time is a better measurement of remaining flying capability. We define $T_{max}$ as maximum flight time, therefore, only flights whose duration is smaller than $T_{max}$ are flyable. Otherwise, we have to divide it into more trips.

%%%%%%%%%--------------------------------------------------------%%%%%%%%%
\subsection{Assumptions}

We assume that at all UAV fly at $v_i$, which is a constant speed of 30km/h.

Since most current UAV on-board battery can only support for around 30-minute flight time, we set $T_{max}$ to be 20 minutes for safe redundancy, which gives a maximum flight distance of 10km. 

%%%%%%%%%%%%%%%%%%%%%%%%%%%%%%%%%%%%%%%%%%%%%%%%%%%%%%%%%%%%%%%%%%%%%%%%%%%%%%%%
\section{Description of Algorithm}

The key of A* algorithm is an evaluation function $f() = g() + h()$, where $g()$ calculates the true path cost, and $h()$ is the heuristic function that estimates the travel time from current node to goal node. We expand $g()$ by adding waiting penalty factor $p$ to $g()$ to use information about node and edge availability. 

%%%%%%%%%--------------------------------------------------------%%%%%%%%%
\begin{algorithm}[H]
\caption{Collision-Free Dynamic Routing}
\begin{algorithmic}[1]
 \renewcommand{\algorithmicrequire}{\textbf{Input:}}
 \renewcommand{\algorithmicensure}{\textbf{Output:}}
 \REQUIRE A non-negative graph $G=(V,E,w)$,\\
    current time $t_{curr}$, no-fly zone $nfz$,
    origin node $v_{ori}$,\\
    destination node $v_{dest}$,
    current schedule $CurrSched$
 \ENSURE  Status code,
    expected travel time $t_{exp}$,\\
    a ordered list of nodes $Path$

 \STATE $CurrSched.updateNFZ(nfz)$
 \STATE initialize OPEN priority queue
 \STATE initialize CLOSED list

 \STATE OPEN.push($v_{ori}$)
 \WHILE {!OPEN.isEmpty()}
   \STATE $q \gets OPEN.pop()$
   \STATE $L \gets G.getSuccessors(q)$
   \FOR {each node $succ_i$ in L}
        \STATE $w \gets CalcCost(succ_i,CurrSched,t_{curr},v_{goal})$
     \IF {$succ_i$ == $v_{dest}$}
        \STATE $t_{exp}, Path \gets backTrack(PARENT)$
        \STATE $CurrSched.update()$
        \RETURN $SUCCESS, Path, t_{exp}$
     \ENDIF
     \STATE OPEN.update()
     \STATE CLOSED.update()
     \STATE OPEN.push($succ_i$,w)
    \ENDFOR
    \STATE CLOSED.add(q)
 \ENDWHILE
 \RETURN $FAILURE$
 \end{algorithmic}
\end{algorithm}

\textit{OPEN} stores the frontier nodes, and \textit{CLOSED} stores the nodes whose successors have been explored. Line 1 to 4 does initialization; exploration of nodes starts from line 5; line 6 and 7 retrieves the current node and its successor node; line 8 generates the successors of the currently expanding nodes, line 9 calculates the cost using algorithm 2; line 10 to 14 checks whether is node is already the goal node and update the schedule accordingly; line 15 to 26 checks whether existing node in CLOSED and OPEN list can have a better upstream node; in the end, the currently explored node will be added to \textit{CLOSED}.
%%%%%%%%%--------------------------------------------------------%%%%%%%%%
\begin{algorithm}[H]
\caption{Calculate Cost}
\begin{algorithmic}[1]
 \renewcommand{\algorithmicrequire}{\textbf{Input:}}
 \renewcommand{\algorithmicensure}{\textbf{Output:}}
 \REQUIRE Successor node $succ$,
    current node $v_{curr}$,
    current schedule $CurrSched$,
    current time $t_{curr}$,
    goal node $v_{goal}$
 \ENSURE  Cost for using this successor $w$
 \STATE $hCost \gets euclideanDistance(t_{curr},v_{goal})
    +edgePen(v_{curr},t_{curr})
    +nodePen(succ,t_{curr})$
 \STATE $gCost  \gets v_{curr}.coSoFar+weight(v_{curr},succ)$
 \RETURN $hCost+gCost$
 \end{algorithmic}
\end{algorithm}

\textit{CalcCost} function is key of the proposed algorithm. The total weight of a path depends on both the time necessary to travel to that node, and waiting penalty along the path (equation \ref{eq:1}).

%%%%%%%%%--------------------------------------------------------%%%%%%%%%
\subsection{Discussion on Optimality}
The proposed algorithm is optimal because the heuristic function for finding weight is admissible. The waiting penalty $P(u,t)$ reflects the actual cost for taking a successor. $edgePen(v_{curr},t_{curr})$ is the waiting time for taking the edge from current node to successor node, and $nodePen(succ,t_{curr})$ is the waiting time when agent reaches the successor node when there are other UAVs hovering on that node. Those two components are unavoidable, thus we view them as actual cost. The rest of the proof can therefore be reduced to the original A* algorithm \cite{c5}.

%%%%%%%%%%%%%%%%%%%%%%%%%%%%%%%%%%%%%%%%%%%%%%%%%%%%%%%%%%%%%%%%%%%%%%%%%%%%%%%%
\section{IMPLEMENTATION}

Now that we have proven the optimality of proposed algorithm via admissibility, we would like to verify our algorithm on a real-life scenario. We select Singapore as a case study for a few reasons: 1) Singapore as a financial center in Southeast Asia is likely to have huge amount of demands for commercial UAV delivery, and 2) Singapore has a high vehicle occupancy rate and relatively small land area. Using UAV for delivery may relieve the demand for ground transport, and therefore reduces traffic congestion.

We implemented a UAV delivery simulation program called Multi-UAV Simulation Engine (MUSE) to test the proposed algorithm. 

%%%%%%%%%--------------------------------------------------------%%%%%%%%%
\subsection{Preparation of Input}

To simulate the UAV traffic, we must have air space structure and delivery requests. 

Under Singaporean setting, we select the rooftop of multistory car parks as landable UAV delivery stations, where there are 77 of them. This is because the rooftop is usually unused in the car park, and each residential area is equipped with a multistory car park. We apply triangulation on the graph to reduce the number of nodes, and manually join collinear nodes on the same direction. The resulting airspace graph is as Fig.1.

  \begin{figure}
      \centering
      \includegraphics[width=8cm]{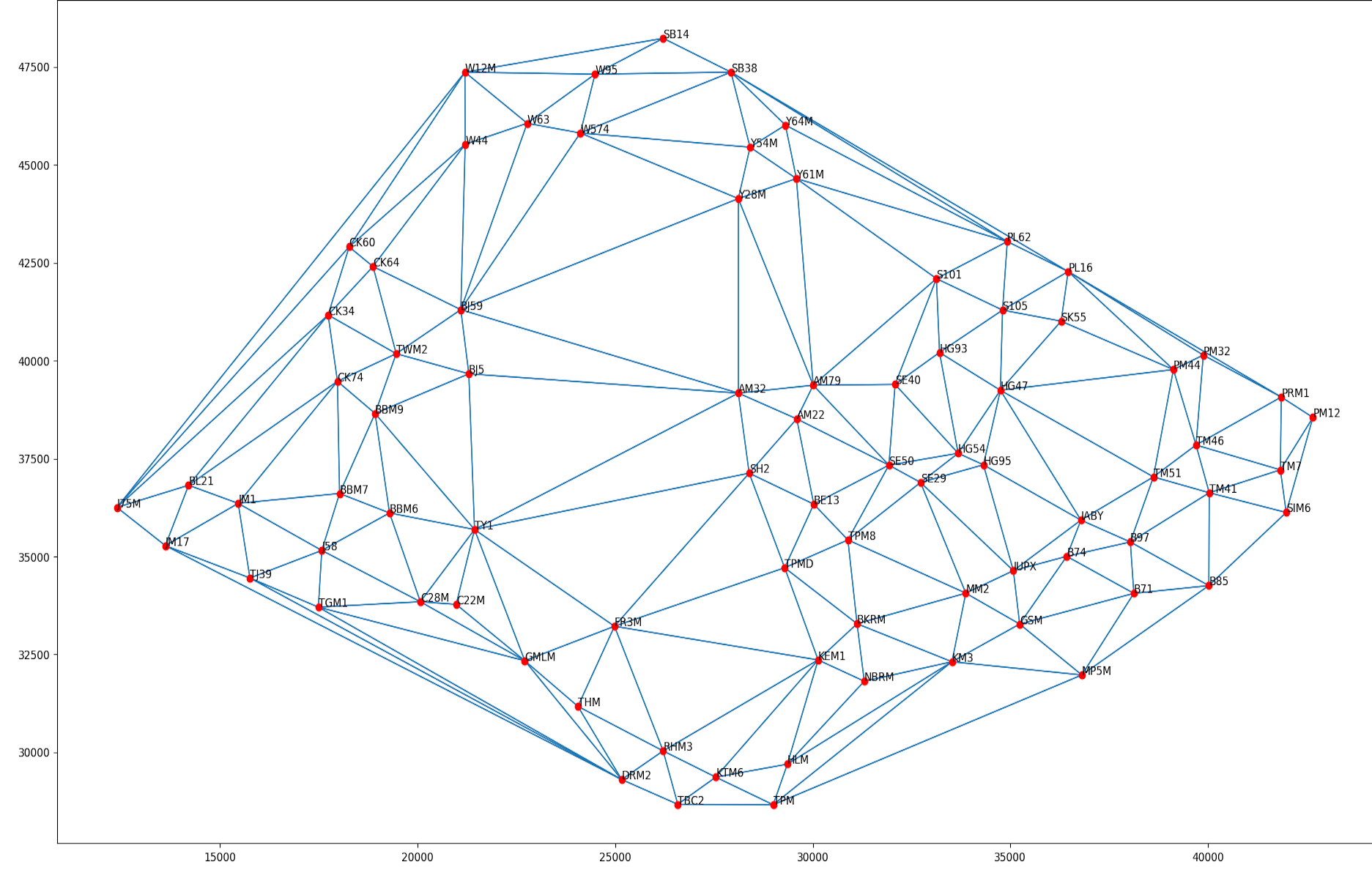}
      \caption{The air space map of Singapore}
      \label{2D_Airspace}
  \end{figure}

Due to a lack of information, we randomly generated the delivery requests with a requested departure time, origin ID, and edge ID.

%%%%%%%%%--------------------------------------------------------%%%%%%%%%
\subsection{Software}
We develop a Java program to simulate the UAV delivery with the proposed algorithm. The program reads airspace structure, request, and no-fly zone schedule. It produces the routing results for each request in JSON format \footnote{The implementation code is open-sourced under MIT license at https://github.com/StevenShi-23/MUSE.}.

%%%%%%%%%--------------------------------------------------------%%%%%%%%%
\subsection{Hardware Setup}
The experiment was conducted on a PC with 2.4 GHz Intel i7 6660U CPU and 16GB memory. 

%%%%%%%%%%%%%%%%%%%%%%%%%%%%%%%%%%%%%%%%%%%%%%%%%%%%%%%%%%%%%%%%%%%%%%%%%%%%%%%%
\section{Analysis of Simulation}
We simulated the UAV delivery from 8:00 am to 8:00 pm, which is 720 minutes. We assume a no-fly zone near Central Business District from 10:00 am to 12:00 am. We simulated from 1000 requests to 10000 requests, and for each number of requests, we repeat the simulation 100 times with different requests to calculate the average. 

%%%%%%%%%--------------------------------------------------------%%%%%%%%%
\subsection{Runtime Analysis}
Fig.2 is a summery of running time in seconds versus number of requests. Notably, it only took our proposed algorithm average 19.45 seconds to route 5000 requests. However, as the number of requests increases, chances of collision increases, thus more nodes need to be expanded and more calculation is done.

  \begin{figure}
      \centering
      \includegraphics[width=8cm]{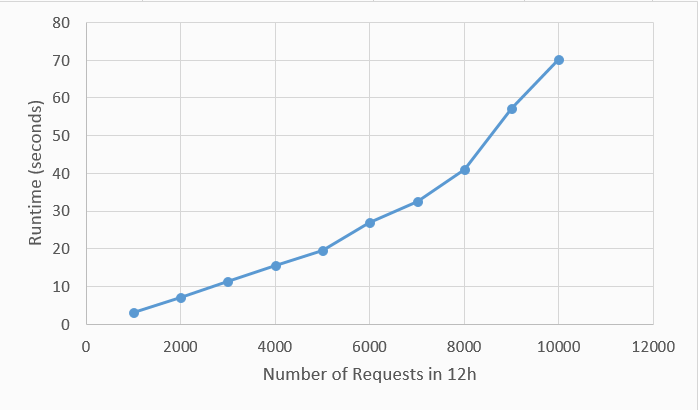}
      \caption{Computational time for finding a path within maximum flight time.}
  \end{figure}

%%%%%%%%%--------------------------------------------------------%%%%%%%%%
\subsection{Success Rate and Average Delay}
We use two metrics to measure the success of our algorithm. Success rate is the percentage of successfully routed requests divided by total number of requests. Failure to route is defined as not being able to find a viable path whose distance is within maximum flight distance. We can see from Fig.3 that the success rate decreases as number of requests increases, and it drops significantly after around 6000 requests. It may come from a few reasons: 1) the distance between requested origin and destination is too far ,or 2) collision avoidance takes too much time for the route to be possible. 
  \begin{figure}
      \centering
      \includegraphics[width=8cm]{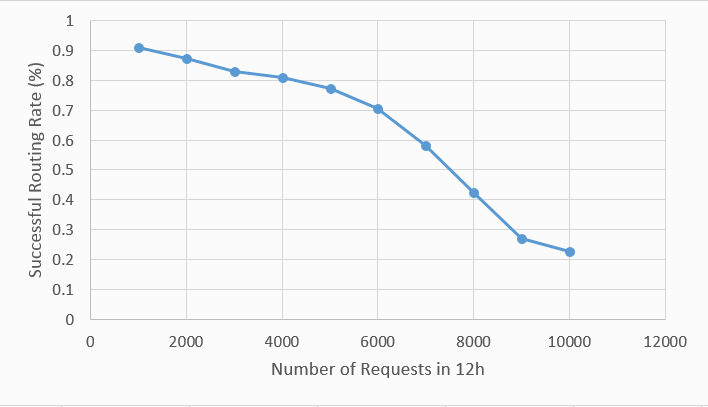}
      \caption{Success rate for finding a flyable path.}
  \end{figure}
  
  Fig.4 shows that as number of requests served per minutes increases, the average delayed time in the air increases. This may verify that 2) is the main cause for the routing failure after 6000 requests.
    \begin{figure}
      \centering
      \includegraphics[width=8cm]{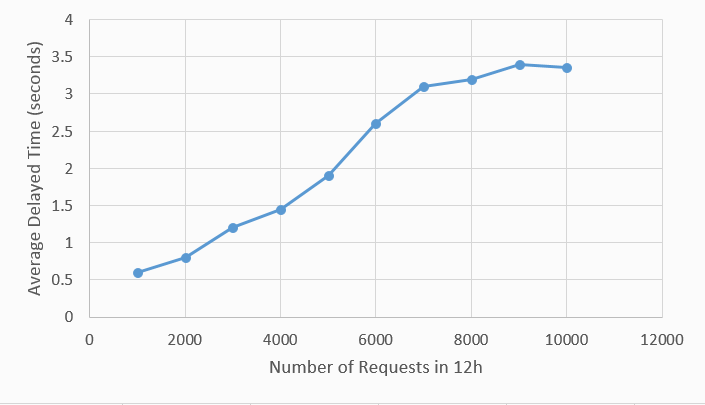}
      \caption{Average delayed time in the air for collision-avoidance and no-fly zone}
  \end{figure}

%%%%%%%%%%%%%%%%%%%%%%%%%%%%%%%%%%%%%%%%%%%%%%%%%%%%%%%%%%%%%%%%%%%%%%%%%%%%%%%%
\section{CONCLUSION}
We introduced an optimal collision-free path planning algorithm based on A* algorithm. The key to obtain waiting penalty is to build an efficient lookup table for all current schedules. The proposed algorithm is capable of dealing with collision-avoidance constraint, no-fly zone constraint, and battery constraint. We also give an implementation for the mentioned algorithm. The runtime for the algorithm scales almost linear to number of requests. It also verifies the intuition that the routing success rate decreases with higher request density. Besides, the success rate drops dramatically at some turning point where there is little space to satisfy new requests.

Although our proposed algorithm is optimal at routing a single new request, the global UAV path planning remains a NP-hard problem \cite{c7}. Future work can be developed on using heuristics to approximate sub-optimal algorithms for global path planning. Also, more realistic requests information could be obtained from authorities to design more fine-grained airspace structure, and study optimal location of UAV delivery stations.

%%%%%%%%%%%%%%%%%%%%%%%%%%%%%%%%%%%%%%%%%%%%%%%%%%%%%%%%%%%%%%%%%%%%%%%%%%%%%%%%
%%%%%%%%%%%%%%%%%%%%%%%%%%%%%%%%%%%%%%%%%%%%%%%%%%%%%%%%%%%%%%%%%%%%%%%%%%%%%%%%

\end{document}